\documentclass[twocolumn,prl,aps,showpacs,amsmath,amssymb,superscriptaddress]{revtex4}

\usepackage{graphics}
\usepackage{epsfig}
\usepackage{amssymb}
\usepackage{amsfonts}

\graphicspath{{figures_eps/}}

\begin{document}

\title{Coherence of Single Electron Sources from Mach-Zehnder Interferometry}

\author{G\'eraldine Haack}
\affiliation{D\'epartement de Physique Th\'eorique, Universit\'e de Gen\`eve, CH-1211 Gen\`eve 4, Switzerland}

\author{Michael Moskalets}
\affiliation{Department of Metal and Semiconductor Physics, NTU "Kharkiv Polytechnic Institute", 61002 Kharkiv, Ukraine}

\author{Janine Splettstoesser}
\affiliation{Institut f\"{u}r Theorie der Statistischen Physik, RWTH Aachen University, D-52056 Aachen, \& Jara - Fundamentals of Future Information Technology, Germany}

\author{Markus B\"uttiker}
\affiliation{D\'epartement de Physique Th\'eorique, Universit\'e de Gen\`eve, CH-1211 Gen\`eve 4, Switzerland}

\date{\today}

\pacs{73.23.Ad, 73.23.-b,72.10.-d}

\begin{abstract}

A new type of electron sources has emerged which permits to inject particles in a controllable manner, one at a time, into an electronic circuit. Such single electron sources make it possible to fully exploit the particles' quantum nature. 
We determine the single-particle coherence length from the decay of the Aharonov-Bohm oscillations as a function of the imbalance of a Mach-Zehnder interferometer connected to a single electron source. The single-particle coherence length is of particular importance as it is an intrinsic property of the source in contrast to the dephasing length.

\end{abstract}
\maketitle

In low temperature electron physics, thus far most electronic transport measurements have used either metallic contacts or superconductors in thermodynamic equilibrium as injectors of electrons. Such sources are suitable to investigate coherent phenomena but are not useful for quantum information as there is no control on the injection time of the single electrons. To the contrary, the recent experimental achievement of single electron sources \cite{Gabelli06,Feve07} heralds a new field in solid-state electronics. 
These single electron sources are high-frequency sources that can emit at a GHz rate single particles with a Lorentzian density profile \cite{OSMB08} of half duration $\Gamma_\mathrm{SES}$.
In the integer Quantum Hall regime, the emitted single particles are injected through a quantum point contact (QPC) into edge states, which are chiral due to suppressed backscattering \cite{Halperin82,Buttiker88} and which take the role of wave guides for electrons. When subject to an appropriate time-periodic voltage, the single electron source (SES) of F\`eve \textit{et al.} \cite{Feve07} can be tuned to inject an electron in the first half cycle and a hole in the second half cycle into the same channel \cite{Mahe10}. Other turnstile like sources separate the two types of carriers into different channels \cite{Battista11} and can produce a stream of electrons only. 
Thanks to their injection tunability, such single-particle state emitters present a powerful potential for quantum information processing \cite{BDV00}: multiparticle exchange, two-particle interference effects and entanglement \cite{OSMB08,SMB09} have already been proposed. However, the appearance of such single-electron sources demands a characterization of their basic property, namely the coherence length $\Lambda$ of the single-particle states they generate \cite{Mandel99}. This length sets the distance over which a single-particle is able to interfere with itself at the output of an interferometer. 

\begin{figure}[ht!]
\centering
\includegraphics[width=8.6cm]{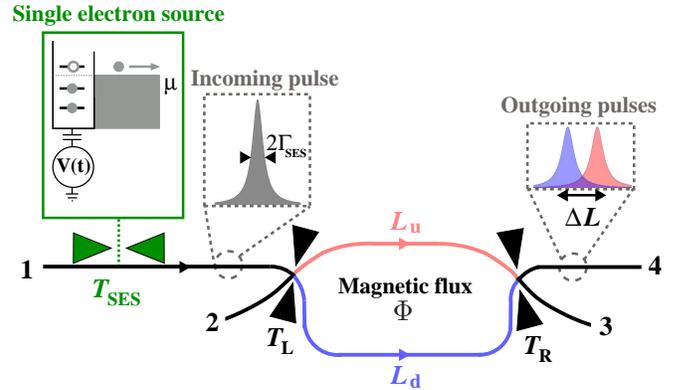}
\caption{(Color online) Single electron source (SES) coupled to a Mach-Zehnder interferometer (MZI) through a quantum point contact (QPC) with transmission probability $T_\mathrm{SES}$. A periodic voltage $V(t) = V(t+ \mathcal{T})$ drives the energy levels of the SES up and down the Fermi energy $\mu$. On average over one time period $\mathcal{T}$, this leads to the injection into the input 1 of the MZI of an electron pulse of half duration $\Gamma_\mathrm{SES}$. The MZI is made of  two QPCs characterized by their transmission probabilities $T_\mathrm{R}$ and $T_\mathrm{L}$. The imbalance of the MZI is defined as $\Delta L = L_\mathrm{u} - L_\mathrm{d}$, with $L_{\mathrm{u}(\mathrm{d})}$ being the length of the upper (lower) arm of the MZI.\label{MZI}}
\end{figure}

In this letter we propose an Aharonov-Bohm (AB) oscillation experiment with an electronic Mach-Zehnder interferometer (MZI) as shown in  Fig.~\ref{MZI} to determine directly this single-particle coherence length $\Lambda$ which reflects the coherence property of the source. In mesoscopics the most natural way to deal with single-particle interference \cite{Imry02} is to measure a current -essentially a single-particle quantity- through multiple-connected structures threaded by a magnetic flux $\Phi$ such as a MZI built in the quantum Hall regime \cite{Ji03}. Due to the AB effect \cite{AB59}, the current measured at the output of the MZI exhibits oscillations as a function of the enclosed magnetic flux $\Phi$ if the size of the interferometer is comparable or smaller than the dephasing length $l_{\varphi}$. In contrast to the single-particle coherence length $\Lambda$, this dephasing length $l_\varphi$ has been widely discussed. It is determined by environment induced decoherence \cite{Roulleau08,NM08} and relaxation processes  \cite{Altimiras10}. In state-of-the-art mesoscopics the size of the MZI is of the order of several microns \cite{Ji03,Litvin07} whereas the dephasing length is of the order of $10\, \mu$m at $20$ mK, thus enabling the observation of AB oscillations.

Due to the finite coherence length of the emitted single-particles, the AB oscillations will be strongly suppressed when the length difference between the upper and lower arms of the interferometer $\Delta L$ exceeds the single-particle coherence length $\Lambda$, see Fig.~\ref{MZI}. In this case the wave packets of both arms can not interfere anymore at the output of the interferometer. By considering the visibility of the AB oscillations defined as the ratio of their amplitude at a given $\Delta L$ to their maximum amplitude at $\Delta L=0$, we have access to the single-particle coherence length. The advantage of such an experiment, compared for example to tomography \cite{degiovanni10}, is that our proposal requires only the measurement of the current and not a more difficult current-correlation measurement. \\

The information we are interested in lies in the initial single-particle state emitted by the source. For this purpose we look at the wave function, $\Psi_{E}(t,x)$ at time $t$ and position $x$ after the SES, see Fig.~\ref{MZI}. It is the product of the plane wave $e^{i(k_E x +\omega t)}$, $k_E$ being the wave number at energy $E=\hbar \omega$, that would have been generated at time $t$ and position $x$ in the absence of the cavity, multiplied by the scattering amplitude $S_\mathrm{SES}$ of the source at the time $t - x/v_\mathrm{D}$ when the wave passed the source:
\begin{equation} 
\Psi_{E}(t,x) = S_\mathrm{SES}\left(t - x/v_\mathrm{D}, E \right) \, e^{i(k_E x +\omega t)} \,.\label{02}
\end{equation}
The scattering amplitude $S_\mathrm{SES}$ introduces a phase in the wave function which has a non-trivial effect on the current \cite{OSMB08,Brouwer}:
\begin{equation}
I_\mathrm{SES}(t,x) \!=\!  \frac{-ie}{2\pi}S_\mathrm{SES}( \!t \!-\! x/v_\mathrm{D},\mu)  \frac{\partial S_\mathrm{SES}^{*}( \!t \!-\! x/v_\mathrm{D},\mu)}{\partial t}\,.
\end{equation}
This current consists of pulses of duration $2\Gamma_\mathrm{SES}$, corresponding to the emission of an electron or to its absorption (corresponding to the emission of a hole) when one level crosses the Fermi energy as indicated in Fig.~\ref{MZI}. During such processes the Fermi sea remains intact \cite{KKL06,Sherkunov} so that it is the scattering amplitude $S_\mathrm{SES}$ at the Fermi energy $\mu$ which describes  the single-particle state emitted by the SES. Therefore the measurement of the current $I_\mathrm{SES}(t,x)$ provides information on the single-particle wave function at a given time $t$ through the scattering amplitude $S_\mathrm{SES}(t\!-\!x/v_\mathrm{D},\mu)$. Such a measurement would require a time-resolution which is fast compared to the temporal variation of the current pulse, exceeding the present experimental possibilities \cite{Gabelli06, Feve07}.\\

To avoid such an experimental constraint, we discuss the determination of the  single-particle coherence length with the help of a MZI built in the quantum Hall regime, see Fig.~\ref{MZI}. Two chiral edge states play the role of one-channel wave-guides for the injected single-particles and two QPCs, characterized by their transmission and reflection probabilities, $T_{\mathrm{L}, \mathrm{R}}$ and $R_{\mathrm{L}, \mathrm{R}} = 1 - T_{\mathrm{L}, \mathrm{R}}$, act as beam splitters for the particles. The lengths of the upper and lower arms are respectively $L_\mathrm{u}$ and $L_\mathrm{d}$ and the imbalance $\Delta L = L_\mathrm{u} - L_\mathrm{d}$ corresponds to the time difference between the traversal of the two paths, $\Delta \tau = \Delta L/v_\mathrm{D}$. Both arms enclose a magnetic flux $\Phi$. The current, measured at contact 4, is calculated at zero temperature with a Floquet scattering matrix approach \cite{Moskalets02}. It is given by the sum of a classical and an interference contributions:
\begin{equation} \label{current}
I_{4}(t,\Phi) = I_{4}^{\mathrm{(cl)}}(t) + I_{4}^{\mathrm{(int)}}(t,\Phi)\, 
\end{equation}
with
\begin{eqnarray}
&&I_{4}^{{\mathrm{(cl)}}}(t) = R_\mathrm{L}R_\mathrm{R} \, I_\mathrm{SES}(t,L_\mathrm{u}) + T_\mathrm{L}T_\mathrm{R}\, I_\mathrm{SES}(t,L_\mathrm{d})  \,, \nonumber \\
&&\!I_{4}^{\mathrm{(int)}}\!(t,\Phi)\! = \!- 2 A I_\mathrm{max} \,{\rm{Im}}\! \left\{ i\,e^{i2\pi \phi} {\cal C}_\mathrm{SES}(t, \Delta \tau) \right\}\,. \nonumber
\end{eqnarray}
We defined the coefficient $A = \sqrt{R_\mathrm{R} R_\mathrm{L} T_\mathrm{R} T_\mathrm{L}}$. 
The interference current $I_{4}^{\mathrm{(int)}}$ depends explicitly on the magnetic flux $\Phi$ and on the imbalance of the MZI through the phase $\phi = \Phi/\Phi_0 + k_\mu\Delta L/(2\pi)$, where $\Phi_0 = h/e$ is the magnetic flux quantum and $k_\mu$ is the wave number at energy $\mu$. Importantly $I_{4}^{\mathrm{(int)}}$ also depends on a factor, namely the single-particle coherence ${\cal C}_\mathrm{SES}(t, \Delta \tau)$, which is directly related to the incoming state through the scattering amplitude $S_\mathrm{SES}$ at the Fermi energy: \begin{equation}
{\cal C}_\mathrm{SES}(t, \Delta \tau) \!=\! \frac{-i\,e}{2\pi I_\mathrm{max}  } \frac{S_\mathrm{SES}(t - \tau_\mathrm{u}, \mu) S_\mathrm{SES}^{*}(t - \tau_\mathrm{d}, \mu) - 1 }{\Delta \tau} \,,
\label{04}
\end{equation}
where $\tau_\mathrm{u,d} = L_\mathrm{u,d}/v_\mathrm{D}$ are the traversal times of the upper and lower arms of the MZI. To obtain a dimensionless measure of the single-particle coherence, we normalized it by the maximum value of the current $I_\mathrm{max}$. In our time-dependent problem, ${\cal C}_\mathrm{SES}(t, \Delta \tau)$ depends on time $t$ and on the time difference $\Delta \tau = \tau_\mathrm{u} - \tau_\mathrm{d}$. When $\Delta \tau$ tends to zero, the single-particle coherence converges to the emitted current $I_\mathrm{SES}/I_\mathrm{max}$, whereas when $\Delta \tau$ exceeds the single-particle coherence time, the coherence ${\cal C}_\mathrm{SES}$ decreases to zero. In this limit the current pulses traveling through different arms have a vanishing overlap at the output as shown in Fig.~\ref{MZI}. Let us note that $\mathcal{C}_\mathrm{SES}(t, \Delta \tau)$ vanishes independently of $\Delta \tau$ at all times $t$ far away from the emission time of an electron $t^-$ or from its absorption time $t^+$ (emission time of a hole), that is far away from the peaks of the current $ \vert I_\mathrm{SES}(t^{\pm})\vert = I_\mathrm{max}$.

Considering a simple model \cite{Feve07} for the SES in which we assume an adiabatic periodic driving potential \cite{MSB08}, the scattering amplitude $S_\mathrm{SES}$ close to the emission time of an electron $t^-$ can be explicitly written \cite{OSMB08} as:
\begin{equation}
S_\mathrm{SES}(t) = \frac{t-t^{-} + i\,\Gamma_\mathrm{SES}}{t -t^{-} - i\,\Gamma_\mathrm{SES}}\,,
\end{equation}
where $\Gamma_\mathrm{SES}  \propto T_\mathrm{SES}/\Omega \sim 10^{-10}$ s is the half duration of the Lorentzian density profile \cite{OSMB08} of the particles emitted by the SES. It depends on the frequency $\Omega = 2\pi/\mathcal{T}$ of the driving voltage $V(t)$ of the SES and on the transmission probability $T_\mathrm{SES}$ of the QPC which couples the SES to the incoming edge state of the MZI (see Fig.~\ref{MZI}). Then at $t = t^{-}$, the single-particle coherence reads: 
\begin{eqnarray}
{\cal C}_\mathrm{SES}(t^{-},\Delta\tau) = \frac{1}{ 1 - i\,\Delta\tau/\Gamma_\mathrm{SES} }\,.
\label{05}
\end{eqnarray}
This expression shows that it is the pulse duration $\Gamma_\mathrm{SES}$ which sets the coherence time. Thus the single-particle coherence length $\Lambda_\mathrm{SES}$ for electrons emitted by the SES is:
\begin{equation}
\label{01}
\Lambda_\mathrm{SES} = v_\mathrm{D}\, \Gamma_\mathrm{SES} \,.
\end{equation}
Here $v_\mathrm{D}$ is the drift velocity of the electrons along the edge of the sample. This velocity ranges from a lower value \cite{Vaart94} of  $10^{5}\,$cm/s up to higher values \cite{McClure09} of the order of $10^{7}$cm/s. Considering the lower limit of this range, we find $\Lambda_\mathrm{SES} \sim 0.1\,\mu$m which is sufficiently small compared to the typical size of a MZI and to the actual dephasing length mentioned above. Therefore, considering the SES as an emitter of single electrons and the electronic MZI as interferometer, it is possible to detect the effect of the single-particle coherence length.\\

Interestingly, Eq.~(\ref{05}) is also valid for a metallic contact with a periodic bias $V(t) = V(t + {\cal T})$, if $V(t)$ is a Lorentzian voltage pulse \cite{KKL06} carrying one electron in one period $\mathcal{T}$. This alternative single-particle source is described by the scattering amplitude $S_{V} = \exp\{ -ie/\hbar \int^{t} dt^{\prime} V(t^{\prime}) \}$ instead of $S_\mathrm{SES}$ and the corresponding emitted current is  $I_{V}(t) = (e^{2}/h)V(t)$ instead of $I_\mathrm{SES}(t)$. For a metallic contact at a dc voltage $V(t) = V$, its single-particle coherence reads: 
\begin{eqnarray}
{\cal C}_\mathrm{dc}(\Delta \tau) = e^{i \pi \Delta\tau/\Gamma_\mathrm{dc}}\, \frac{\sin\left( \pi\Delta\tau/\Gamma_\mathrm{dc} \right)}{\pi\Delta\tau/\Gamma_\mathrm{dc}}\,,
\label{07}
\end{eqnarray}
with $\Gamma_\mathrm{dc} = \Lambda_\mathrm{dc}/v_D= h/(eV)$. This coherence reflects the structure of a wave-packet incident in a single quantum channel subject to a dc-bias. If carriers are injected through a tunnel contact with transmission probability $T$, the mean time between carriers transmitted through the MZI, $h/(eVT)$, increases. However as long as the energy dependence of the transmission $T$ can be neglected, the coherence function is independent of $T$: it is given by Eq.~(\ref{07}) as for the case of a perfect quantum channel. \\

\begin{figure}[ht!]
\centering
\includegraphics[width=6.5cm]{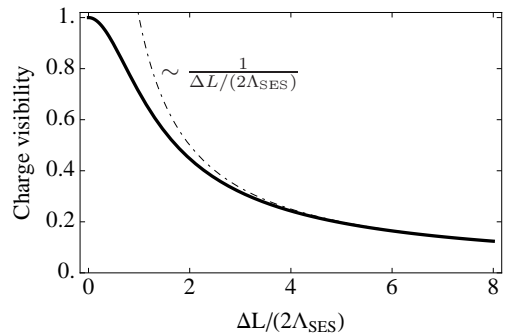}
\caption{Charge visibility $\nu_Q$ as a function of the imbalance $\Delta L/(2\Lambda_\mathrm{SES})$ (thick line). At small imbalance, $\nu_Q$ decays slower than $1/\Delta L$ (dashed line).\label{charge_visi}}
\end{figure}

The full reconstruction of the single-particle coherence ${\cal C}_\mathrm{SES}(t,\Delta\tau)$ requires a time-resolved measurement of the interference current $I_{4}^{(\mathrm{int})}$ (see Eq.~(\ref{current})). However the real advantage of our setup is that much simpler time-averaged measurements are sufficient to determine the single-particle coherence length. By integrating the current $I_4(t,\Phi)$ over one half-period (during which an electron is emitted), we find a charge with visibility $\nu_Q$ equal to the absolute value of the coherence:
\begin{equation} \label{charge_visibility}
\nu_Q = \left\vert {\cal C}_\mathrm{SES}(t^{-},\Delta \tau/2)  \right\vert = \frac{1}{\sqrt{1 + (\Delta L/(2\Lambda_\mathrm{SES}))^2}}\,.
\end{equation}
This charge visibility decays proportional to $1/\Delta L$  with increasing interferometer imbalance as shown in Fig.~\ref{charge_visi}. Experimentally it is necessary to distinguish a decrease of the visibility due to the finite coherence length of the source from a decrease in visibility due to dephasing \cite{Ji03}. To keep the dephasing constant, the total length of the arms of the interferometer has to be kept constant. It is thus favorable to have a detectable effect of the imbalance $\Delta L$ on the visibility even when $\Delta L$ is small.
\begin{figure}[ht!]
\centering
\includegraphics[width=8.6cm]{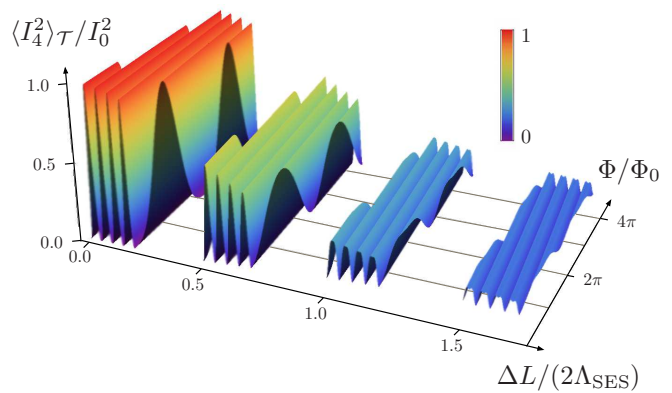}
\caption{(Color online) Mean squared current in units of $I_0^2 = e^2 \Omega/(2\pi^2 \Gamma_\mathrm{SES})$ as a function of the magnetic flux $\Phi$ and the imbalance $\Delta L$, $T_\mathrm{R}= T_\mathrm{L}=0.5$. For clarity we show only 4 cuts of the function at $\Delta L/(2\Lambda_\mathrm{SES}) = 0.0, \,0.5, \,1.0$ and $1.5$., allowing to see the projection for different constant values of $\Delta L/(2\Lambda_\mathrm{SES})$. 
\label{fig2}}
\end{figure}

A quantity that exhibits a stronger variation with $\Delta L$ is the mean squared current $\left\langle I_{4}^{2} \right\rangle_{\mathcal{T}} = \int_{0}^{\cal T}  \left(I_{4}(t,\Phi)\right)^{2}\, dt/{\cal T}$. This time-averaged quantity exhibits AB oscillations as a function of the magnetic flux with two periods $\Phi_{0}$ (first harmonic) and $\Phi_{0}/2$ (second harmonic), see Fig.~\ref{fig2}. The AB oscillations are maximum when the imbalance is zero (complete overlap in contact $4$ of the particle pulses traveling in different arms), whereas they strongly decrease when the imbalance becomes larger than the single-particle coherence length, $\Delta L/(2\Lambda_\mathrm{SES}) \gtrsim 1.5$. At larger imbalance, the function tends to a constant value $(R_\mathrm{R}R_\mathrm{L})^2 + (T_\mathrm{R}T_\mathrm{L})^2$ corresponding to the classical contribution of the mean squared current. When $\Delta $ exceeds $L \sim 2\Lambda_\mathrm{SES}$, the second harmonic starts contributing. As for the current $I_4(t,\Phi)$, it is $\Gamma_\mathrm{SES}$ which sets the vanishing of the AB oscillations and thus gives information on the single-particle coherence length.
\begin{figure}[ht!]
\centering
\includegraphics[width=7cm]{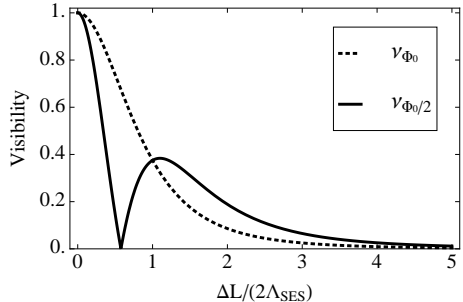}
\caption{Visibilities of the first harmonic $\nu_{\Phi_0}$ (dashed curve), and of the second harmonic, $\nu_{\Phi_{0}/2}$ as a function of the imbalance $\Delta L/(2\Lambda_\mathrm{SES})$.\label{fig3}}
\end{figure}
To characterize the decay of the oscillations for the first and second harmonics, we define a visibility for each harmonic, $\nu_{\Phi_0}$ and $\nu_{\Phi_{0}/2}$, as the ratio of the amplitude of oscillations at a given interferometer imbalance $\Delta L$ to the maximum amplitude of oscillations, i.e., the one of the balanced interferometer characterized by $\Delta L = 0$:
\begin{eqnarray} \label{08}
\nu_{\Phi_0} &=& \frac{1}{\left(1+ \left(\frac{\Delta L}{2 \Lambda_\mathrm{SES}}\right)^{2} \right)^2 }\,, \\
\nu_{\Phi_0/2} &=& \frac{\left| 1 - 3\left(\frac{\Delta L}{2 \Lambda_\mathrm{SES}}\right)^{2} \right|}{\left(1+ \left(\frac{\Delta L}{2 \Lambda_\mathrm{SES}}\right)^{2} \right)^3 } \,.
\end{eqnarray}
These visibilities are much more sensitive functions of the MZI's imbalance $\Delta L$ than the charge visibility as showed in Fig.~\ref{fig3}. This makes the measurement of $\Lambda_\mathrm{SES}$ easier. The strong decay of the visibilities of the AB oscillations of the mean squared current can be explained by the electron-hole asymmetry inherent to any mesoscopic circuit with energy-dependent transmission. Consequently electron and hole contributions partially cancel each other in the interference part, leading to the predicted strong decay. \\

To summarize, the realization of novel single electron sources demands a characterization of their coherence properties. We have demonstrated that coupling such a source to a MZI and measuring the mean squared current permits to determine the coherence length of single electrons. Thus a current measurement that is slow compared to the evolution of the single electron pulses is nevertheless sufficient to extract information on the quantum nature of the emitted states. The coherence of single electron states is an essential concept for electronics and also for engineering few electron quantum states. 

This work is supported by the Swiss NSF, the Swiss NCCR on Quantum Science and Technology, 
(QSIT) and the European Network NanoCTM. J. S. also thanks the Ministry of Innovation NRW.



\begin{thebibliography}{1}

\bibitem{Gabelli06}
J. Gabelli \textit{et al.}, Science, {\bf 313}, 499 (2006).

\bibitem{Feve07}
G. F\`{e}ve \textit{et al.}, Science {\bf 316}, 1169 (2007).

\bibitem{OSMB08} 
S. Ol'khovskaya, J. Splettstoesser, M. Moskalets, and M. B\"uttiker, Phys. Rev. Lett. {\bf 101}, 166802 (2008).

\bibitem{Halperin82} 
B. I. Halperin, Phys. Rev. B {\bf25}, 2185 (1982)

\bibitem{Buttiker88}
M. B\"uttiker, Phys. Rev. Lett. {\bf 38}, 9375 (1988).


\bibitem{Mahe10} 
A. Mah\'e \textit{et al.}, Phys. Rev. B {\bf82}, 201309 (2010);
M. Albert, C. Flindt  and M. B\"uttiker, Phys. Rev. B {\bf82}, 041407 (2010).


\bibitem{Battista11}
F. Battista and P. Samuelsson,
arXiv:1006.0136 (2010).


\bibitem{BDV00}
C. H. Bennett and D. P. DiVincenzo, Nature (London) {\bf 404}, 247 (2000).


\bibitem{SMB09}
J. Splettstoesser, M. Moskalets and M. B\"uttiker, Phys. Rev. Lett. {\bf 103}, 076804 (2009).

\bibitem{Mandel99}
L. Mandel, Rev. Mod. Phys. {\bf 71}, S274 (1999).

\bibitem{Imry02} Y. Imry, \textit{Introduction to Mesoscopic Physics}, Oxford University Press, USA, (2002).

\bibitem{Ji03}
Y. Ji \textit{et al.}, Nature {\bf 422}, 415 (2003).


\bibitem{AB59}
Y. Aharonov and D. Bohm, Phys. Rev. {\bf 115}, 485 (1959).

\bibitem{Roulleau08}
P. Roulleau \textit{et al.}, Phys. Rev. Lett. {\bf 100}, 126802 (2008);
E. Bieri \textit{et al.}, Phys. Rev. B {\bf79}, 245324 (2009).

\bibitem{NM08}
C. Neuenhahn and F. Marquardt, New J. Phys. {\bf 10}, 115018 (2008);


\bibitem{Altimiras10}
C. Altimiras \textit{et al.}, Nat. Physics {\bf 6}, 34 (2010);
H. le Sueur \textit{et al.}, Phys. Rev. Lett. {\bf 105}, 056803 (2010).

\bibitem{Litvin07}
L. V. Litvin, H.-P. Tranitz, W. Wegscheider and C. Strunk, Phys. Rev. B. {\bf 75}, 033315 (2007).


\bibitem{degiovanni10} 
C. Grenier \textit{et al.}, arXiv:1010.2166 (2010).
 
\bibitem{Brouwer} P. W. Brouwer, \textit{Phys. Rev. B} {\bf 58}, 10135 (1998);
J. E. Avron \textit{et al.}, \textit{Phys. Rev. B} {\bf 62}, R10618 (2000);
M. B\"{u}ttiker, H. Thomas and A. Pr\^{e}tre, Z. Phys. B {\bf 94}, 133 (1994).

 
\bibitem{KKL06}
J. Keeling, I. Klich and L. S. Levitov, Phys. Rev. Lett. {\bf 97}, 116403 (2006).

\bibitem{Sherkunov}
Y. Sherkunov, J. Zhang, N. d'Ambrumenil and B. Muzykantskii, Phys. Rev. B {\bf 80}, 041313 (2009).

\bibitem{Moskalets02}
M. Moskalets and M. B\"uttiker, Phys. Rev. B {\bf 66}, 205320 (2002). 


\bibitem{MSB08}
M. Moskalets, P. Samuelsson and M. B\"uttiker, Phys. Rev. Lett. {\bf 100}, 086601 (2008).

\bibitem{Vaart94}
N. C. Van der Vaart \textit{et al.}, Phys. Rev. Lett. {\bf 73}, 320 (1994).

\bibitem{McClure09} 
D. T. McClure \textit{et al.}, Phys. Rev. Lett. {\bf103}, 206806 (2009).




\end{thebibliography}
\end{document}